\documentclass[fleqn,twoside,superscriptaddress,preprintnumbers,amsmath,amssymb]{article}
\usepackage[headings]{espcrc2}

\readRCS
$Id: espcrc2.tex,v 1.2 2004/02/24 11:22:11 spepping Exp $
\usepackage{graphicx}
\usepackage[figuresright]{rotating}

\newcommand{\AmS}{{\protect\the\textfont2
  A\kern-.1667em\lower.5ex\hbox{M}\kern-.125emS}}
\def\kt{{k_{_\perp}}}
\def\cO#1{{{\cal{O}}}\left(#1\right)}
\newcommand{\lqcd}{\Lambda_{_{_{\rm QCD}}}}
\newcommand{\dd}{{\rm d} }

\title{Inclusive hadronic distributions in jets and
sub-jets with jet axis from color current}

\author{Redamy P\'erez-Ramos\address[MCSD]{II. Institut f\"ur Theoretische Physik,
Universit\"at Hamburg, \\ 
        Luruper Chaussee 149
22761 Hamburg,
Germany}%
}

\runtitle{Inclusive hadronic distributions in jets and sub-jets with jet axis from color current}
\runauthor{R. P\'erez-Ramos}

\begin{document}

\begin{abstract}
The hadronic $\kt$-spectrum and the gluon to quark average multiplicity
ratio $r=N_g/N_q$ inside a high energy
sub-jet are determined from a precise definition of the jet axis,
including corrections of relative magnitude
 ${\cal O}(\sqrt{\alpha_s})$ with respect to the Modified Leading Logarithmic
Approximation (MLLA),  in the limiting spectrum approximation
(assuming an infrared cut-off $Q_0 =\lqcd$).
The  results for the $\kt$-spectrum in the limiting spectrum approximation
are found to be, after normalization, in
impressive agreement with measurements by the CDF
collaboration and the ratio has not been measured yet.
\vspace{1pc}
\end{abstract}

\maketitle

\section{Introduction}
\label{section:intro}

The collimation of hadrons in jets is a basic phenomenon of high energy
collisions and its quantitative understanding is an important task for QCD.
Simple differential characteristics of a jet are their energy and
multiplicity angular profiles. 
The collimation of energy and multiplicity in the jet
follows from the dominance of gluon bremsstrahlung processes in the parton
cascade evolution. Whereas the former is more sensitive to the hard 
processes inside a jet the second one is sensitive to the soft parton emissions
from the primary parton. 

The characteristics of soft particle production, such as particle
multiplicities, inclusive distributions and correlation functions, are
derived in QCD in the Double Logarithmic Approximation (DLA)
and the Modified Leading Logarithmic Approximation (MLLA) (for
review, see \cite{Basics}), which takes into account the leading double
logarithmic terms and the single logarithmic corrections. These azimuthally
averaged quantities can be obtained from an evolution equation for the
generating functional of the parton cascade. This equation provides also
Next-to-MLLA (NMLLA) corrections taking into account energy conservation of parton
splittings with increased accuracy. The corresponding hadronic observables
can be obtained using the concept of Local Parton Hadron Duality (LPHD)
\cite{LPHD}, which has turned out a successful description of many hadronic
phenomena (see, e.g. \cite{KhozeOchs}).

The inclusive $\kt$-distribution of particles
inside a sub-jet with respect to the jet axis 
has been computed at MLLA accuracy
in the limiting spectrum
approximation~\cite{PerezMachet}, {\it i.e.} assuming an infrared cutoff $Q_0$ equal to
$\lqcd$ ($\lambda\equiv\ln( Q_0/\lqcd)= 0$) (for a review, see also \cite{Basics}).
MLLA corrections, of relative magnitude $\cO{\sqrt{\alpha_s}}$, 
were shown to be quite
substantial ~\cite{PerezMachet}. Therefore, corrections of order 
$\cO{\alpha_s}$, that is NMLLA, have been incorporated 
in \cite{PAM,PRL}.

Another characteristic prediction of 
QCD is the asymptotic increase of the
mean average particle multiplicity in a gluon
jet over the quark jet by the ratio of color factors
$r=N_g/N_q \to N_c/C_F=9/4$, which is also modified by the similar set of 
corrections ${\cal O}(\sqrt{\alpha_s})$ \cite{MultTheory,MALAZA,DREMIN}.
In particular, this analysis has been extended to multiparticle 
production inside sub-jets with a precise prescription of the 
jet axis, which has been, for both observables, identified with
the direction of the energy flux \cite{ochsperez}.

The starting point of this analysis is the MLLA Master Equation for
the {\em Generating Functional} (GF) $Z=Z(u)$
of QCD jets \cite{Basics}, where $u=u(k)$ is a certain probing
function and $k$, the four-momentum of the outgoing parton.
Together with the initial condition at threshold, the GF determines the
jet properties at all energies. 
Within this logic, the leading 
(DLA, ${\cal O}(\sqrt{\alpha_s})$) 
and next-to-leading (MLLA, ${\cal O}(\alpha_s)$) approximations are complete.
The next terms (NMLLA, ${\cal O}(\alpha_s^{3/2})$) 
are not complete but they include an important
contribution which takes into account energy conservation and an
improved behavior near threshold. Indeed, some results for such NMLLA terms have been
studied previously for global observables and have been found to
better account for recoil effects and to drastically affect
multi-particle production \cite{DokKNO,CuypersTesima}. 

The main results
of this work have been published in \cite{PAM,PRL,ochsperez}.
Experimentally, the CDF collaboration at the Tevatron reported on
$\kt$-distributions of unidentified charged hadrons in jets produced in
$p\bar{p}$ collisions at $\sqrt{s}=1.96$~TeV~\cite{CDF}.

\section{MLLA evolution equations}
\label{section:sis}

\subsection{Inclusive spectrum}

\begin{figure}[ht]
\begin{center}
  \includegraphics[height=3.2cm]{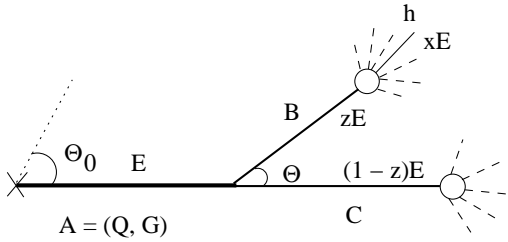}
\end{center}
\caption{Parton A with energy $E$ splits into parton B
(respectively C) with energy $zE$ (respectively, $(1-z)E$)
which fragments into a hadron $h$ with energy $xE$.}
\label{fig:spplit}
\end{figure}

We start by writing the MLLA evolution equations for the fragmentation
function $D_{B}^h\left(x\big/z, zE\Theta_0, Q_0\right)$ of
a parton B (energy $zE$ and transverse momentum $\kt=zE\Theta_0$)
into a gluon (represented by a hadron $h$ (energy $xE$)
according to LPHD \cite{LPHD}) 
inside a jet A of energy $E$ for the process depicted
in Fig.~\ref{fig:spplit}.
As a consequence of angular ordering in parton cascading,
partonic distributions inside a quark and gluon jet,
$Q,G(z)=x\big/z\ D_{Q,G}^h\left(x\big/z, zE\Theta_0, Q_0\right)$, obey
the system of two coupled  equations \cite{RPR2} (the subscript ${}_y$ denotes
$\partial/\partial{y}$) following from the MLLA master equation \cite{Basics}
\begin{eqnarray}
 Q_{y}\!\!&\!\!=\!\!&\!\!\int_0^1\!\! \dd z  \frac{\alpha_s}{\pi}\Phi_q^g(z)
\bigg[\Big( Q(1-z)-Q\Big) +  G(z) \bigg]\label{eq:qpr}\\
 G_{y}\!\!&\!\!=\!\!&\!\!\int_0^1 \dd z \frac{\alpha_s}{\pi}
\bigg[\Phi_g^g(z) \Big( G(z) + G(1-z)\Big.\cr
\!\!&\!\!-\!\!&\!\!\Big.G\Big)+n_f \Phi_g^q(z)\Big(2 Q(z)- G\Big)\bigg].
\label{eq:gpr}
\end{eqnarray}
$\Phi_A^B(z)$ denotes the Dokshitzer-Gribov-Lipatov-Altarelli-Parisi (DGLAP) 
splitting functions \cite{Basics},
\begin{eqnarray}
\Phi_q^g(z)\!\!&\!\!=\!\!&\!\!C_F\left(\frac2z+\phi_q^g(z)\right),\cr
\Phi_g^g(z)&\!=&\!T_R\left[z^2+(1-z)^2\right],\cr
\Phi_g^g(z)\!\!&\!\!=\!\!&\!\!2N_c\left(\frac1z+\phi_g^g(z)\right),
\end{eqnarray}
where $\phi_q^g(z)=z-2$ and $\phi_g^g(z)=(z-1)(2-z(1-z))$ are regular
as $z\to0$, 
$C_F=(N_c^2-1)/2N_c$, $T_R=n_f/2$ ($N_c=3$ is the number of colors
for $SU(3)_c$ and
$n_f=3$ is the number of light flavors we consider).
The running coupling of QCD ($\alpha_s$) is given by
$$
\alpha_s=\frac{2\pi}{4N_c\beta_0(\ell+y+\lambda)},
\,4N_c\beta_0=\frac{11}{3}N_c-\frac43T_R
$$ 
and
$$
\ell=\left(1/x\right)\ ,\ y=\ln\left(\kt\big/Q_0\right)\ ,
\ \lambda=\ln\big(Q_0/\lqcd\big),
$$
where $Q_0$ is the collinear cut-off parameter. Moreover,
\begin{eqnarray}
G & \equiv & G(1) = xD_G^h(x,E\Theta_0,Q_0),\cr
Q & \equiv & Q(1) = xD_Q^h(x,E\Theta_0,Q_0).
\end{eqnarray}
At small $x\ll z$, the fragmentation functions behave as
$$
B(z)\approx \rho_B^{h} \left(\ln\frac{z}{x},\ln\frac{zE\Theta_0}{Q_0}\right)
=\rho_B^h\left(\ln z + \ell, y\right),
$$
$\rho_B^{h}$ being a slowly varying function of two logarithmic
variables $\ln (z/x)$ and $y$ that describes the ``hump-backed''
plateau~\cite{HBP}. Since recoil effects should be largest in hard
parton splittings, the strategy followed in this work is
to perform Taylor expansions (first advocated for in \cite{Dremin})
of the non-singular parts of the integrands
in~(\ref{eq:qpr},\ref{eq:gpr}) in powers of
$\ln z$ and $\ln(1-z)$, both considered  small  with respect to
$\ell$ in the hard splitting region  $z\sim 1-z =\cO{1}$
%
\begin{eqnarray}
\label{eq:logic1}
B(z) = B(1) + B_\ell(1) \ln z + \cO{\ln^2z}\,
\end{eqnarray}
with $z\leftrightarrow 1-z$.
Each $\ell$-derivative giving an extra $\sqrt{\alpha_s}$ factor
(see~\cite{RPR2}), the terms $B_\ell(1) \ln z$ and $B_\ell(1)
\ln\left(1-z\right)$ yield NMLLA corrections to the solutions of
(\ref{eq:gpr}).
Making used of (\ref{eq:logic1}), after integrating over 
the regular parts of the DGLAP splitting functions, 
while keeping the singular
terms unchanged, one gets after some algebra
($\gamma_0^2=2N_c \alpha_s/\pi$)~\cite{PAM,PRL}
\begin{eqnarray}\label{eq:solq}
Q(\ell,y)\!\!&\!\!=\!\!&\!\!\delta(\ell)+\frac{C_F}{N_c}\!\!
\int_0^{\ell}\! \dd \ell'\!\int_0^{y}\! \dd y' \gamma_0^2(\ell'+y')\times\\
&&\hskip -1.5cm\Big[1-\tilde a_1\delta(\ell'-\ell)+\tilde a_2\delta(\ell'-\ell)
\psi_\ell(\ell',y')\Big]G(\ell',y'),\cr
G(\ell,y)\!\!&\!\!=\!\!&\!\!\delta(\ell)
+\int_0^{\ell}\! \dd \ell'\!\int_0^{y}\! \dd y' \gamma_0^2(\ell'+y')
\times\cr
&&\hskip -1.5cm\Big[1 - a_1\delta(\ell'-\ell)
a_2\delta(\ell'-\ell)\psi_\ell(\ell',y')\Big]G(\ell',y'),\label{eq:solg}
\end{eqnarray}
where the constants take the values $a_1\approx0.935$, $\tilde a_1=3/4$,
$a_2\approx0.08$ and $\tilde a_2\approx0.42$ for $n_f=3$.
Defining $F(\ell,y)=\gamma_0^2(\ell+y)G(\ell,y)$, we can exactly solve 
the self-contained equation (\ref{eq:solg}) by performing the inverse Mellin
transform:
\begin{equation}\label{eq:mellin}
F(\ell,y)=\int\frac{d\omega d\nu}{(2\pi i)^2}e^{\omega\ell}e^{\nu y}
{\cal F}(\omega,\nu), 
\end{equation}
such that the NMLLA solution reads
\begin{eqnarray}
G(\ell,y)\!&\!=\!&\!(\ell+y+\lambda)\!\!\int\!\!\frac{d\omega d\nu}{(2\pi i)^2}
e^{\omega\ell}e^{\nu y}\!\int_0^{\infty}\!\!\!\frac{\dd s}{\nu+s}\cr
&&\hskip -1cm\times\left(\frac{\omega
\left(\nu+s\right)}{\left(\omega+s\right)\nu}\right)^{\sigma_0}
\left(\frac{\nu}{\nu+s}\right)^
{\sigma_1+\sigma_2}e^{-\sigma_3\,s},
\label{eq:mellinnmlla}
\end{eqnarray}
where 
$$
\sigma_0=\frac{1}{\beta_0(\omega-\nu)},\quad
\sigma_1=\frac{a_1}{\beta_0},
$$
and 
$$
\sigma_2=-\frac{a_2}{\beta_0}(\omega-\nu),\quad 
\sigma_3=-\frac{a_2}{\beta_0}+\lambda.
$$
As can be seen, the NMLLA coefficient $a_2$ is very small.
Therefore, the NMLLA solution (\ref{eq:mellinnmlla}) of 
(\ref{eq:solg}) can be approximated
by the MLLA solution of $G(\ell,y)$ ({\it i.e.} taking $a_2\approx0$),
which is used in the following to compute the inclusive $\kt$-distribution. 
As demonstrated
in \cite{RPR2}, taking the limits $a_2\approx0$ and $\lambda\approx0$ in (\ref{eq:mellinnmlla}), the integral representation can be reduced to the 
known MLLA formula, which can be written in terms of Bessel series
in the limiting spectrum approximation \cite{Basics}.
To get a quantitative idea on the difference between MLLA and NMLLA
gluon inclusive spectrum, the reader is reported
to the appendix B of \cite{PAM}.
The magnitude of $\tilde a_2$, however, indicates that the NMLLA
corrections to the inclusive quark jet spectrum may not be negligible
and should be taken into account. 
After solving (\ref{eq:solg}), the solution of (\ref{eq:solq}) reads
\begin{eqnarray}
Q(\ell,y)\!\!&\!\!=\!\!&\!\!\frac{C_F}{N_c}\left[G(\ell,y)
+\Big(a_1-\tilde a_1\Big)G_\ell(\ell,y)\right.\cr
\!\!&\!\!+\!\!&\!\!\left.
\left(a_1\Big(a_1-\tilde a_1\Big)+\tilde a_2-a_2\right)G_{\ell\ell}(\ell,y)
\right].\label{eq:ratioqg}
\end{eqnarray}
It differs from the MLLA expression given in \cite{PerezMachet} by the term
proportional to $G_{\ell\ell}$, which can be deduced from the subtraction of 
$(C_F/N_c)\times$(\ref{eq:solg}) to Eq.~(\ref{eq:solq}).

\subsection{Particle mean average multiplicity}

Integrating the system (\ref{eq:qpr},\ref{eq:gpr}) over the energy 
fraction $x$ leads the average multiplicity of particles
produced in the jet:
$$
N_A(Y_{\Theta_0})=\int_0^1 dx D_A(x,E\Theta_0,Q_0).
$$
where $Y_{\Theta_0}\equiv\ell+y=\ln(E\Theta_0/Q_0)$. Therefore, 
an equivalent system of two-coupled evolution 
equations can be deduced and expanded, following the same
logic that led to (\ref{eq:solq},\ref{eq:solg}). However,
in this case, the expansion in $\sqrt\alpha_s$
is performed for $Y_\Theta\gg\ln z\sim\ln(1-z)$, where similarly,
$z\sim1$. We limit ourselves here to the NMLLA expression of $r$ that reads
\begin{equation}\label{eq:QGratio}
r=\frac{N_g}{N_q}=r_0(1-r_1\gamma_0-r_2\gamma_0^2)+{\cal O}(\gamma_0^3),
\end{equation} 
where the asymptotic value of $r$ is $r_0=N_c/C_F=9/4$. 
The MLLA term $r_1$ has been calculated in \cite{MultTheory}.
The coefficients $r_k$ can be
obtained from the Taylor expansions
of $N_A^h(Y_\Theta+\ln x)$ and $N_A^h(Y_\Theta+\ln (1-x))$
for $Y_\Theta\gg\ln x$ and $Y_\Theta\gg\ln(1-x)$ in the evolution
equations \cite{Dremin}.
The values of $r_k$ for $n_f=3$ 
read $r_1=0.185$ and $r_2=0.426$ \cite{Dremin}. 
The NMLLA solution
for the mean multiplicity in a gluon jet 
is found as \cite{Dremin}
\begin{eqnarray}
N_g^h(Y_\Theta)\!\!&\!\!\simeq\!\!&\!\! {\cal K}\left(Y_\Theta\right)^{-c_1/\beta_0}\cr
\!\!&\!\!\times\!\!&\!\!\exp{\left(\frac{2}{\sqrt{\beta_0}}\sqrt{Y_\Theta}-
\frac{2c_2}{\beta_0^{3/2}\sqrt{Y_\Theta}}\right)}\label{eq:NGhNMLLA}
\end{eqnarray}
with $c_1=0.28$, $c_2=0.38$ for 
$n_f=3$ and ${\cal K}$, the LPHD
normalization factor.
The pre-exponential term $(Y_\Theta)^{-c_1/\beta_0}$
is the MLLA contribution to $N_g^h$, while
the one $\propto c_2$, the NMLLA one.
\section{Single inclusive 
$k_\perp-$distribution
of charged hadrons and the gluon to quark mean average multiplicity
ratio $r$ at NMLLA}

\subsection{Inclusive $k_\perp-$distribution}

\begin{figure}[h]
\begin{center}
\includegraphics[height=2.3cm]{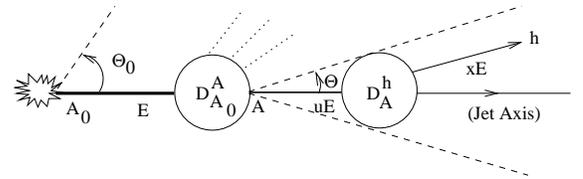}
\caption{\label{fig:distri} Inclusive production of hadron $h$ at angle
$\Theta$ inside a high energy jet of total opening angle
$\Theta_0$ and energy $E$.}
\end{center}
\end{figure}

Computing the single inclusive $k_\perp-$ distribution
and the ratio $r=N_g/N_q$ inside a sub-jet requires
the definition of the jet axis. The starting 
point of our approach consists in considering
the correlation between two particles
(h1) and (h2) of energies $E_1$ and $E_2$ which form a
relative angle $\Theta$ inside one jet of total 
opening angle $\Theta_0>\Theta$ \cite{DDT}. 
Weighting over the energy $E_2$ of particle (h2), 
this relation leads to the correlation
between the particle (h=h1)
and the energy flux, which we identify
with the jet axis (see Fig.~\ref{fig:distri}) \cite{PerezMachet}. Thus,
the correlation and the relative transverse momentum 
$k_\perp$ between (h1) and (h2) are replaced by the correlation,
and transverse momentum of (h1) with
respect to the direction of the energy flux. 
Finally, we obtain
the double differential
spectrum $\dd^2N/\dd{x}\,\dd\Theta$ of a hadron produced  with energy
$E_1=xE$ at angle
$\Theta$ (or $k_\perp\approx xE\Theta$) with respect to the jet axis.
As demonstrated in~\cite{PerezMachet}, the correlation reads
\begin{equation}
\frac{\dd^2N}{\dd x\,\dd\ln{\Theta}}=
\frac{\dd}{\dd\ln\Theta}F_{A_0}^{h}\left(x,\Theta,E,\Theta_0\right),
\label{eq:DD}
\end{equation}
where $F_{A_0}^{h}$ is given by the convolution
of two fragmentation functions
\vbox{
\begin{eqnarray}
F_{A_0}^{h} &\equiv&
\sum_{A=g,q}\int_x^1 \dd u D_{A_0}^A\left(u,E\Theta_0,uE\Theta\right)\cr
&&\hskip 1.5cm \times D_A^{h}\left(\frac{x} {u},uE\Theta,Q_0\right),
\label{eq:F}
\end{eqnarray}
}
$u$ being the energy fraction of the intermediate parton $A$.
$D_{A_0}^A$ describes
the probability to emit $A$ with energy $uE$ off the parton $A_0$
(which initiates the jet),  taking into account the evolution
of the jet between $\Theta_0$ and $\Theta$. $D_{A}^h$ describes the
probability to produce the hadron $h$ off $A$ with energy fraction $x/u$ and
transverse momentum $\kt\approx uE\Theta\geq Q_0$
(see Fig.~\ref{fig:distri}).
As discussed in \cite{PerezMachet}, the convolution
(\ref{eq:F}) is dominated by $u\sim 1$ and therefore
$D_{A_0}^A\left(u,E\Theta_0,uE\Theta\right)$ is determined by the DGLAP evolution
\cite{Basics}. On the contrary, the distribution
$D_A^{h}\left(\frac{x}{u},uE\Theta,Q_0\right)$ 
at low $x\ll u$ reduces to the hump-backed
plateau,
\begin{equation}\label{eq:expansion1}
\tilde D_A^h(\ell + \ln u,y) \stackrel{x\ll u}
{\approx} \rho_A^{h}(\ell + \ln u, Y_\Theta + \ln u),
\end{equation}
with $Y_\Theta=\ell+y=\ln (E\Theta/Q_0)$. Performing the Taylor expansion of
$\tilde D$ to the second order in $(\ln u)$ and plugging it
into Eq.~(\ref{eq:F}) leads to 
\begin{equation}
\label{eq:coldef}
F_{g, q}^{h}(\ell,y) = \frac{1}{N_c}\langle C\rangle_{g, q}(\ell,y)\ G(\ell,y).
\end{equation}
where $\langle C\rangle_{g, q}$ can be written in the symbolic way
\begin{equation}
\label{eq:Fdev}
\langle C\rangle_{g, q}(\ell,y)
\simeq f_0(\ell,y)+f_1(\ell,y)\sqrt\alpha_s+f_2(\ell,y)\alpha_s,
\end{equation}
where $f_k$ follow from the DGLAP evolution equations \cite{Basics}
and can be symbolically written as
\begin{equation}\label{eq:fk}
f_k\simeq
\int_x^1 \dd u\, u (\ln u)^{k} D_{A_0}^A\left(u,E\Theta_0,uE\Theta\right)
\end{equation}
and corrections ${\cal O}(\alpha_s^{n/2})$ from the $n-$logarithmic derivatives of
the inclusive spectrum:
$$
\frac{1}{D_A}\frac{\partial^nD_A}{\partial\,
\ell^n}\simeq{\cal O}(\alpha_s^{n/2}).
$$
The exact formul{\ae} of the color currents are
reported in \cite{PerezMachet,PAM,PRL}.
Indeed, since soft particles are less sensible to the energy balance,
the correlation (\ref{eq:F}) disappears for these particles, 
leading to the sequence of factorized terms written in (\ref{eq:Fdev}).
The first two terms in Eq.~(\ref{eq:Fdev}) correspond
to the MLLA distribution calculated in \cite{PerezMachet}
when  $\tilde D_A^{h}$ is evaluated at NLO and its derivative at LO.
NMLLA corrections arise from their respective calculation at
NNLO and NLO, and, mainly in practice, from  the third term, which was
exactly computed in \cite{PAM,PRL}.
Indeed, since $x/u$ is small, the inclusive spectrum
$\tilde D_A^h(\ell,y)$ with $A=G,Q$ are given by 
the solutions (\ref{eq:mellinnmlla}) and (\ref{eq:ratioqg})
of the next-to-MLLA evolution equations (\ref{eq:solq})
and (\ref{eq:solg}) respectively. However, because of the smallness
of the coefficient $a_2$, $G(\ell,y)$ shows no significant
difference from MLLA to NMLLA. As a consequence, we use the MLLA
expression (\ref{eq:mellinnmlla}) for $G(\ell,y)$ with $a_2=0$, and the NMLLA
(\ref{eq:ratioqg}) for $Q(\ell,y)$.
The functions $F_{g}^{h}$ and $F_{q}^{h}$ are related to the gluon
distribution {\it via} the color currents $\langle C\rangle_{g, q}$
which can be seen as the average color charge carried
by the parton $A$ due to the DGLAP evolution from $A_0$ to $A$.

\begin{figure}
\begin{center}
\includegraphics[height=5cm,width=6cm]{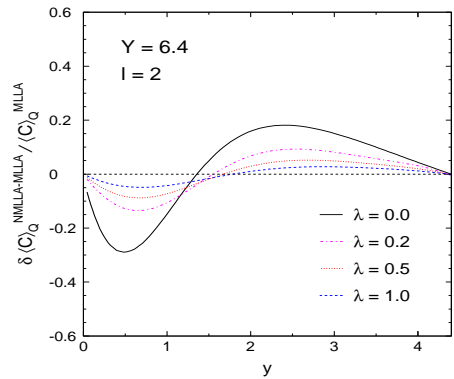}
\caption{\label{fig:CCQlambda}NMLLA corrections to the color current of
a quark jet with $Y_{\Theta_0}=6.4$ and $\ell=2$ for various
values of $\lambda$.}
\end{center}
\end{figure}

This calculation has also been extended beyond the limiting spectrum,
$\lambda\ne 0$,  to take into account hadronization effects in
the production of ``massive'' hadrons, $m=\cO{Q_0}$~\cite{finitelambda}.
We used, accordingly, the more general 
MLLA solution of (\ref{eq:solg}) with $a_2=0$ 
for an arbitrary $\lambda\ne0$,
which can only be performed numerically.
The NMLLA (normalized) corrections to the MLLA result are displayed in
Fig.~\ref{fig:CCQlambda} for different values 
$\lambda=0,0.5,1$.
It clearly indicates that
the larger $\lambda$, the smaller the NMLLA corrections.
In particular, they  can be as large as $30\%$ at the limiting
spectrum ($\lambda=0$) but no more than $10\%$ for $\lambda=0.5$.
This is not surprising since $\lambda\ne 0$ ($Q_0\ne\lqcd$) reduces
the parton emission in the infrared sector and, thus, higher-order
corrections.

\begin{figure}
\begin{center}
\includegraphics[height=7.5cm,width=7.5cm]{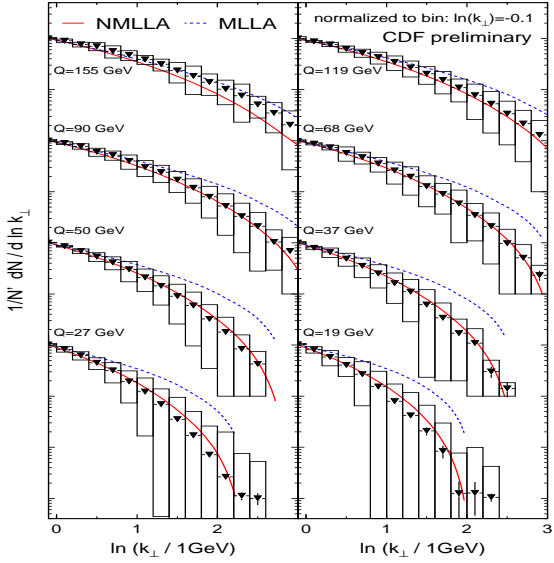}
\caption{\label{fig:CDF-NMLLA}CDF preliminary results for the inclusive
$\kt$ distribution at various hardness $Q$ in comparison to MLLA and
NMLLA predictions at the limiting spectrum ($Q_0=\Lambda_{QCD}$); the boxes are
the systematic errors.}
\end{center}
\end{figure}

The double differential spectrum $(\dd^2N/\dd\ell\,\dd{y})$
can now be determined from the NMLLA color currents
(\ref{eq:Fdev}) by using the quark and gluon distributions.
Integrating over $\ell$ leads to the single inclusive $y$-distribution 
(or $\kt$-distribution) of hadrons inside a quark or a gluon jet:
\begin{equation}
\left(\frac{\dd N}{\dd y}\right)_{A_0}
= \int_{\ell_{\rm min}}^{Y_{\Theta_0}-y}\;\dd\ell
\;\left(\frac{\dd^2N}{\dd\ell\, \dd y}\right)_{A_0}.
\label{eq:lmin}
\end{equation}
The MLLA framework does not specify down to which values of
$\ell$ (up to which values of $x$) the double differential
spectrum $(\dd^2N/\dd\ell\,\dd{y})$
should be integrated over. Since $(\dd^2N/\dd\ell\,\dd{y})$ becomes
negative (non-physical) at small values of $\ell$
(see e.g.~\cite{PerezMachet}), we chose the lower bound $\ell_{\rm min}$
so as to guarantee the positiveness of $(\dd^2N/\dd\ell\,\dd{y})$
over the whole $\ell_{\rm min}\le \ell \le Y_{\Theta_0}$ range
(in practice, $\ell_{\rm min}^g\sim 1$ and $\ell_{\rm min}^q\sim 2$). 
Having successfully computed the single $\kt$-spectra including NMLLA
corrections, we now compare the result with existing data.
The CDF collaboration at the Tevatron
recently reported on preliminary measurements  over a wide
range of jet hardness, $Q=E\Theta_0$, in $p\bar{p}$ collisions at
$\sqrt{s}=1.96$~TeV~\cite{CDF}. CDF data, including systematic errors,
are plotted in
Fig.~\ref{fig:CDF-NMLLA} together with the MLLA predictions of
\cite{PerezMachet} and the present NMLLA calculations, both
at the limiting spectrum ($\lambda=0$) and taking $\lqcd=250$~MeV.
The experimental distributions suffering from large normalization errors,
data and theory are normalized to the same bin, $\ln(k_\perp/1\,GeV)=-0.1$.
The agreement between the CDF results and the NMLLA distributions over
the whole $\kt$-range is particularly good.
In contrast, the MLLA predictions prove reliable in a much smaller
$\kt$ interval. At fixed jet hardness (and thus $Y_{\Theta_0}$),
NMLLA calculations prove
accordingly trustable in a much larger $x$ interval.

\begin{figure}
\begin{center}
\includegraphics[height=5cm, width=6cm]{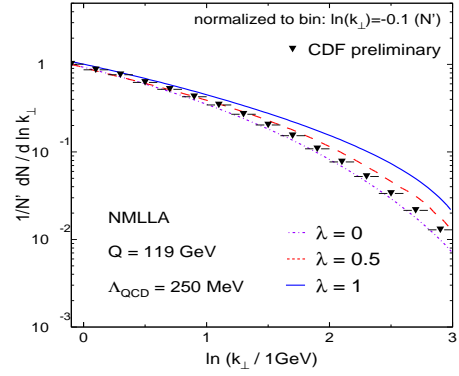}
\caption{\label{fig:CDF-NMLLAlambda}CDF preliminary results ($Q=119$~GeV)
for inclusive $\kt$
distribution compared with NMLLA predictions beyond the limiting spectrum.}
\end{center}
\end{figure}

Finally, the kt-distribution is determined with respect to the jet energy
flow (which includes a summation over secondary hadrons in
energy-energy correlations). In experiments, instead, the jet axis
is determined exclusively from all particles inside the jet.
The question of the matching of these two definitions goes beyond
the scope of this letter.
The NMLLA $\kt$-spectrum has also been calculated beyond the limiting
spectrum, by plugging (\ref{eq:mellinnmlla}) with $a_2=0$ 
into (\ref{eq:lmin}), as 
illustrated in Fig.~\ref{fig:CDF-NMLLAlambda}.
However, the best description of CDF preliminary data is reached at the limiting
spectrum, or at least for small values of $\lambda< 0.5$, which is not too
surprising since these inclusive measurements 
mostly involve pions. Identifying produced hadrons would offer the
interesting possibility to check a dependence
of the shape of $\kt$-distributions on the hadron species,
such as the one predicted in Fig.~\ref{fig:CDF-NMLLAlambda}.
Moreover, the softening of 
the $\kt$-spectra with increasing hadron masses predicted in 
Fig.~\ref{fig:CDF-NMLLAlambda} 
is an observable worth to be measured, as this would provide an 
additional and independent check of the LPHD hypothesis beyond the limiting spectrum. 
This could only be achieved if the various species of hadrons inside a jet can be identified experimentally. Fortunately, it is likely to be the case at the LHC, where the ALICE~\cite{identificationalice}
and CMS~\cite{identificationcms} experiments at the Large Hadron Collider have good 
identification capabilities at not too large transverse momenta.

\subsection{Ratio $r=N_g/N_q$}

Integrating (\ref{eq:F}) over the energy fraction $x$ yields the
corresponding sub-jet multiplicity $\hat N_{A_0}^h$ of hadrons  
inside the angular range $\Theta<\Theta_0$ of the jet $A_0$ \cite{ochsperez}
\begin{eqnarray}
\hat N_{A_0}^h(\Theta;E,\Theta_0)
\!\!\!&\!\!\!\approx\!\!\!&\!\!\!\sum_{A=q,g}\int_{u_0}^1 du\,u\,
D_{A_0}^A\left(u,E\Theta_0,uE\Theta\right)\cr
&&\hskip 1cm\times N_A^{h}\left(uE\Theta,Q_0\right),\label{eq:Nconv}
\end{eqnarray}
where $N_A^h$ is the number of hadrons (partons at scale $Q_0$)
produced inside the sub-jet $A$
of total virtuality $uE\Theta\geq Q_0$ and $u_0=Q_0/E\Theta$. 
Within the leading parton approximation,
we expand the multiplicity
$N_A^h(uE\Theta,Q_0)$ in (\ref{eq:Nconv}) at $u\sim 1$. 
Similarly to the logic applied in (\ref{eq:expansion1}) for
the computation of the $k_\perp-$distribution,
we take the logarithmic dependence of the average
multiplicity $N_A^h$, such that for
$\ln u\ll Y_\Theta\equiv\ln(E\Theta/Q_0)$ and 
$E\Theta\gg \Lambda$,
one obtains the average multiplicity 
$\hat N_{A_0}^h(Y_{\Theta_0},Y_{\Theta})$
of soft hadrons 
within the sub-jet of opening angle $ \Theta$ with respect to the
energy flow
\begin{equation}\label{eq:Ncc}
\hat N_{q,g}^h(Y_{\Theta_0},Y_{\Theta})
\approx\frac{1}{N_c}\langle C\rangle _{q,g}(Y_{\Theta_0},Y_{\Theta})\,
N_g^h(Y_{\Theta}),
\end{equation}
where $N_g^h(Y_{\Theta})$ is written in  
(\ref{eq:NGhNMLLA}) and
$\langle C\rangle _{q,g}(Y_{\Theta_0},Y_{\Theta})$
is the average color current of partons 
forming the energy flux. We write the color current as
$$
\langle C\rangle _{q,g}(Y_{\Theta_0},Y_{\Theta})
\simeq\tilde f_0+
\tilde f_1\sqrt\alpha_s+
\tilde f_2\alpha_s,
$$
where $\tilde f_k\equiv\tilde f_k(Y_{\Theta_0},Y_{\Theta})$ 
follow from (\ref{eq:QGratio}) and (\ref{eq:fk})
and corrections ${\cal O}(\alpha_s^{n/2})$ from the logarithmic 
derivatives of the average multiplicity (\ref{eq:NGhNMLLA}):
$$
\frac{1}{N_A}\frac{d^nN_A}{dY_\Theta^n}\simeq\alpha_s^{n/2}.
$$
The exact formul{\ae} of $\langle C\rangle _{q,g}(Y_{\Theta_0},Y_{\Theta})$
are given in the appendix of \cite{ochsperez}.
The ratio of the gluon to the quark jet average multiplicity reads
$$
\frac{\hat N_g^h}{\hat N_q^h}(Y_{\Theta_0},Y_{\Theta})=
\frac{\langle C\rangle _{g}(Y_{\Theta_0},Y_{\Theta})}
{\langle C\rangle _{q}(Y_{\Theta_0},Y_{\Theta})}.
$$
For this quantity, in the limit
$\Theta\to\Theta_0$, the appropriate ratio (\ref{eq:QGratio})
$r=\frac{N_g}{N_q}=r_0(1-r_1\gamma_0-r_2\gamma_0^2$)
is recovered.

\begin{figure}[h]
\begin{center}
\includegraphics[height=5cm, width=6cm]{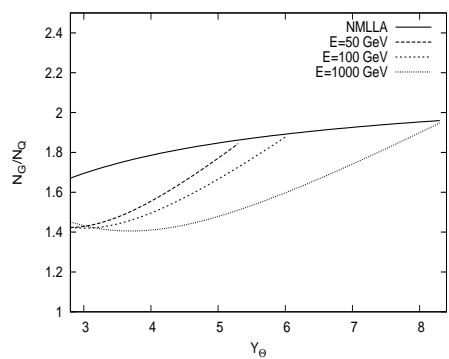}
\caption{\label{fig:MultRat}Multiplicity ratio $N_g/N_q$ for jets and sub-jets
as function of energy variable
$Y=\ln(E\Theta/Q_0)$.}
\end{center}
\end{figure}

Next we study the consequences for the behavior of the ratio
at full angles $\Theta_0\sim 1$ and for sub-jets at reduced angles $\Theta$.
Results for the ratio using our formulae for the color currents 
$\langle C\rangle _{q,g}$ \cite{ochsperez} are depicted
in Fig.\,\ref{fig:MultRat}.
The full line shows the ratio (\ref{eq:QGratio}) 
for an isolated $A_0$ jet as a function of 
$Y_{\Theta_0}=\ln(E\Theta_0/Q_0)$.
The three curves below correspond to the sub-jet 
energies $E=50,100,1000$ GeV and the intersection
points with (\ref{eq:QGratio}) to the limit $\Theta\to\Theta_0$.
Therefore, the mixing of quark and
gluon jets reduces the multiplicity 
in the gluon jet and increases it in
the quark jet at sufficiently high energies and
the effect from intermediate processes in 
the new definition of multiplicity ratio
amounts to about 20\% at a reduced energy scale $Y_\Theta\sim Y_{\Theta_0}-2$. 

\section{Conclusion}

We have studied a definition of the jet which arises from the 2-particle
correlation. The jet axis corresponds to the energy weighted direction
of particles in a given cone $\Theta_0$. In this way smaller angles (or transverse
momenta) can be meaningfully determined.

The new effect from 2-particle correlations is the appearance of the
``color current'' in the result, which reflects the possibility of
intermediate quark-gluon processes. We then studied the consequences
of this analysis on the measurement of differential ($k_\perp$-distribution) 
and global (particle multiplicity) observables, where
we have added MLLA and NMLLA corrections to the known LO results \cite{Basics}.

The single inclusive $\kt$-spectra is
determined within this logic and the agreement between NMLLA predictions
and CDF data in $p\bar{p}$ collisions at the Tevatron \cite{CDF} is
impressive, indicating very small overall
non-perturbative corrections and giving further support
to LPHD \cite{LPHD}. 

The main for the ratio
$r=N_g/N_q$ reflects the possibility of
intermediate quark-gluon processes. This effect can be seen by the scale
breaking between jets and sub-jets at the same scale $E\Theta$ but different
energies $E$ and opening angles $\Theta$. 
Typical effects are of the order of 20\%. However, these effects
on this observable have not
been measured yet.

\end{document}